\documentclass[preprint,12pt,amssymb,floatfix,amsmath]{elsarticle}
\usepackage[usenames]{color}
\usepackage{subfigure}

 \usepackage{graphicx}

\usepackage{amssymb}
\usepackage{bm}

\journal{Physica C}

\def\BSCCO{Bi$_2$Sr$_2$CaCu$_2$O$_{8+\delta}$}
\newcommand{\dx}{\nabla_x}
\newcommand{\dy}{\nabla_y}

\newcommand{\di}{\nabla_i}

\newcommand{\df}{\delta O_N}
\newcommand{\dfd}{\delta O^{(0)}_N}
\newcommand{\fz}{\langle O_n\rangle}

\newcommand{\bcoup}{{\alpha}}
\newcommand{\dcoup}{{\beta}}
\newcommand{\gcoup}{{\gamma}}

\begin{document}

\begin{frontmatter}



\title{Electronic Liquid Crystal Physics of Underdoped Cuprates
}


\author[EK]{Eun-Ah Kim}
\author[ML]{Michael J. Lawler}
\address[EK]{Cornell University}
\address[ML]{Binghamton University}

\begin{abstract}
Recent observations of broken symmetries have partly demystified the pseudogap phase. Here we review evidence for long-range intra-unit-cell(IUC) nematic order and its unexpectedly strong coupling to the phase of the fluctuating stripes in the pseudogap states of underdoped \BSCCO. In particular, we focus on the analysis techniques that reveal this evidence in scanning tunneling spectroscopy data, the definition of the extracted IUC nematic order parameter, and a phenomenological theory of the coupling between the IUC nematic order and the previously reported coexisting fluctuating stripes.    
We also present a microscopic mechanism of IUC nematic order driven by on-site and near-neighbor repulsions. Finally we discuss open questions in the context of these results. 
\end{abstract}


\end{frontmatter}

\section{Introduction}
\label{sec:intro}
The wide variety of systems discussed in this special issue form strong empirical evidence that electronic liquid crystals generically emerge out of correlated electronic systems in a quantum regime. On the one hand, the uncertainty principle challenges against a single sweep numerical solution of a fermionic hamiltonian, when neither the single particle kinetic energy nor the inter-particle interaction energy can be ignored. On the other hand, the conflict between kinetic energy and interaction energy indeed appear to find a compromise in electronic liquid crystals\cite{nematic-review}, as first proposed in Ref.~\cite{Kivelson:1998nature}. This observation invites what we call "middle-up/down approach" to quantum phenomena of correlated systems: seeking insights in experimental data from the perspective of symmetries (middle$\rightarrow$down) and then feeding those insights into theory (middle$\rightarrow$up). What guides this approach are symmetry principles.

\begin{figure}
\centering
\includegraphics[height=.25\textheight]{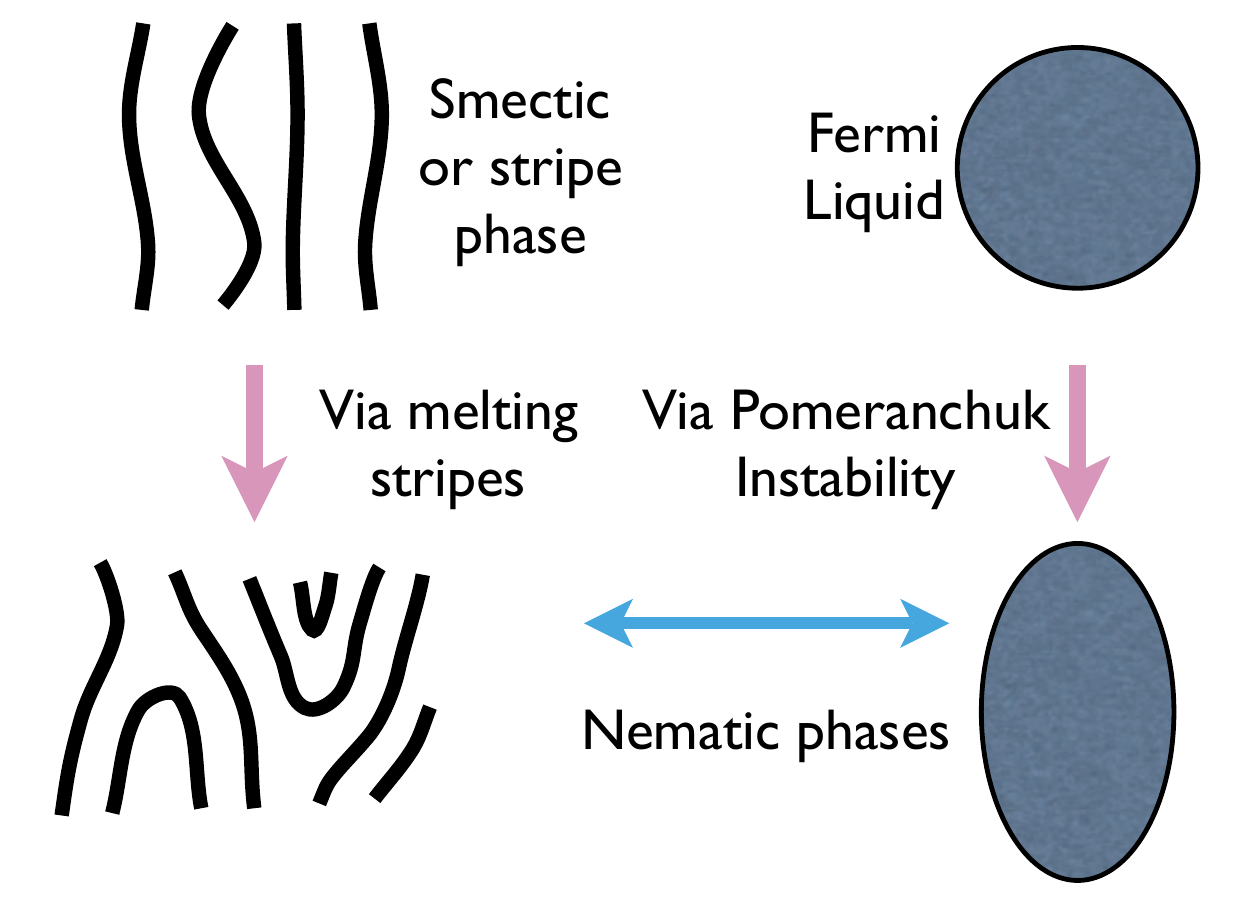}
\caption{Two perspectives on electronic liquid crystals. Orientational order can arise as a shape instability of the Fermi surface (right). Conversely, it could arise from the melting of a unidirectional wave (stripe) pattern through the proliferation of dislocations. The resulting phases in each case can be adiabatically connected to each other.}
\label{fig:strong-weak}
\end{figure}

From symmetry principles, electronic nematic and smectic phases are analogues of liquid crystalline nematic and smectic phases. When a collection of anisotropic molecules called ``nematogens'' are in a liquid phase, the system is invariant under infinitesimal translations and rotations. In a nematic phase, the system still has this translational symmetry but the rotational symmetry is broken and the system is only symmetric under a rotations by $180^\circ$.  In a smectic phase, the translational symmetry is also reduced in one of spatial directions and a modulated density forms in that direction. This modulation automatically breaks rotational symmetry in space as well. Electronic nematic and smectic phases share similar symmetries. The smectic phase would be a kind of unidirectional charge density wave while a nematic phase could be viewed in two ways, as the melting of the unidirectional waves or as a shape instability of the Fermi surface (see Fig. \ref{fig:strong-weak}). However, the order parameters of these phases is different from their classical counter parts because the highest symmetry they may have is not that of an isotropic space but that of, for example, a square lattice\cite{nematic-review}. 

However, electronic liquid crystals as electronic phases in the quantum regime face new challenges and possibilities. One new challenge is in figuring out a mechanism for nematic formation without a pre-formed ``nematogen''. In liquid crystal nematics the prolonged shape of the ``nematogen'' molecule provides an entropy driven mechanism: at low temperatures it is easier for them to move around if they all point in the same direction. For electronic nematics, a mechanism for spatial symmetry breaking is more subtle though an extended range interaction is one possibility. 
One new phenomena arising in the electronic version of liquid crystals are the quantum phase transitions that exist between them. While phase transitions between different liquid crystalline phases are determined as a balance between energy and entropy at finite temperature, transitions between different electronic liquid crystal phases can be controlled even at zero temperature by quantum fluctuations and can lead to novel non-Fermi liquid physics. 

\begin{figure}
\centering
\includegraphics[height=.25\textheight]{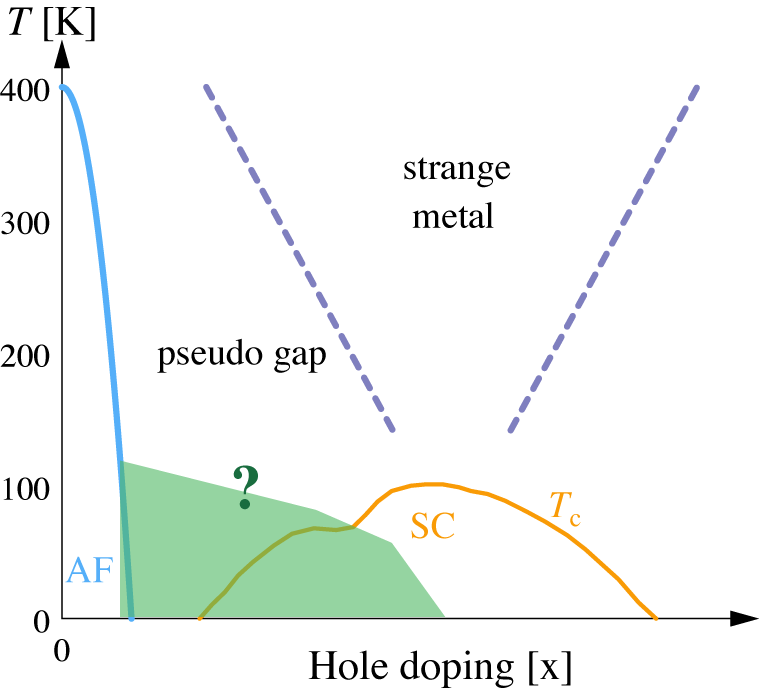}
\caption{A schematic phase diagram for cuprates.}
\label{fig:pd}
\end{figure}

In this article, we review our recent progress in understanding electronic liquid crystal physics using a "middle-up/down approach" to underdoped cuprates.
Cuprates are paradigmatic strongly correlated systems whose electronic properties change dramatically as one scans through the phase diagram Fig.~\ref{fig:pd}. In the underdoped region below a doping- and probe-dependent temperature scale  $T^*$, cuprates exhibit a loss of low energy states below some energy scale: a ``pseudogap''. Whether this region involves a spontaneous symmetry breaking has been a topic of fierce debate ever since the discovery of cuprates. At the simplest level, this debate has been waiting for evidence of symmetry breaking. Remarkably, such evidence has started to accumulate recently\cite{fauque:2006,Daou:2010,hinkov:2008,Howald19082003,Lawler:2010fk} and here we will focus on evidence of symmetry breaking towards the formation of an electronic liquid crystal in \BSCCO.  

The rest of the article is organized as follows. In section \ref{sec:nematic}, we will discuss an intra-unit-cell (IUC) nematic order parameter and a smectic order parameter as defined and measured in Ref~\cite{Lawler:2010fk}. 
In section \ref{sec:coupling}, a proposal for an effective theory describing the coupling between the two order parameters and its validity as tested in Ref~\cite{Mesaros22072011} are reviewed.
In section \ref{sec:emery}, a microscopic mechanism for IUC nematic and other IUC symmetry breaking, based on inter-site repulsion in the Emery model for CuO$_2$ plane~\cite{PhysRevB.84.144502} is discussed. We close the article in section \ref{sec:closing} with a discussion of recent theoretical developments on the subject and interesting future directions.

\section{Intra-Unit-Cell Electronic Nematic and Fluctuating Smectic}
\label{sec:nematic}
\begin{figure}[t]
   \includegraphics[height=.15\textheight]{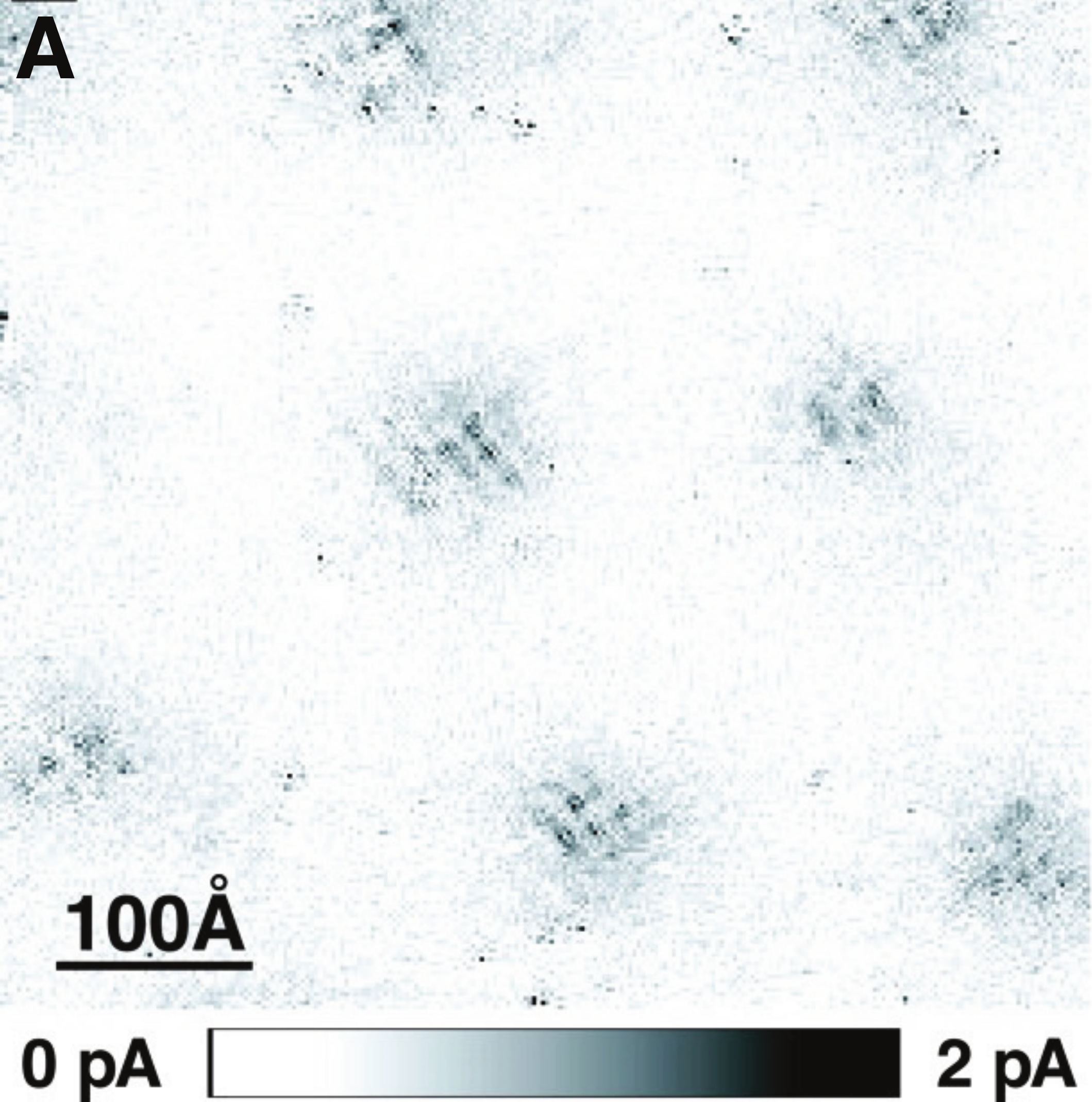}
   \includegraphics[height=.15\textheight]{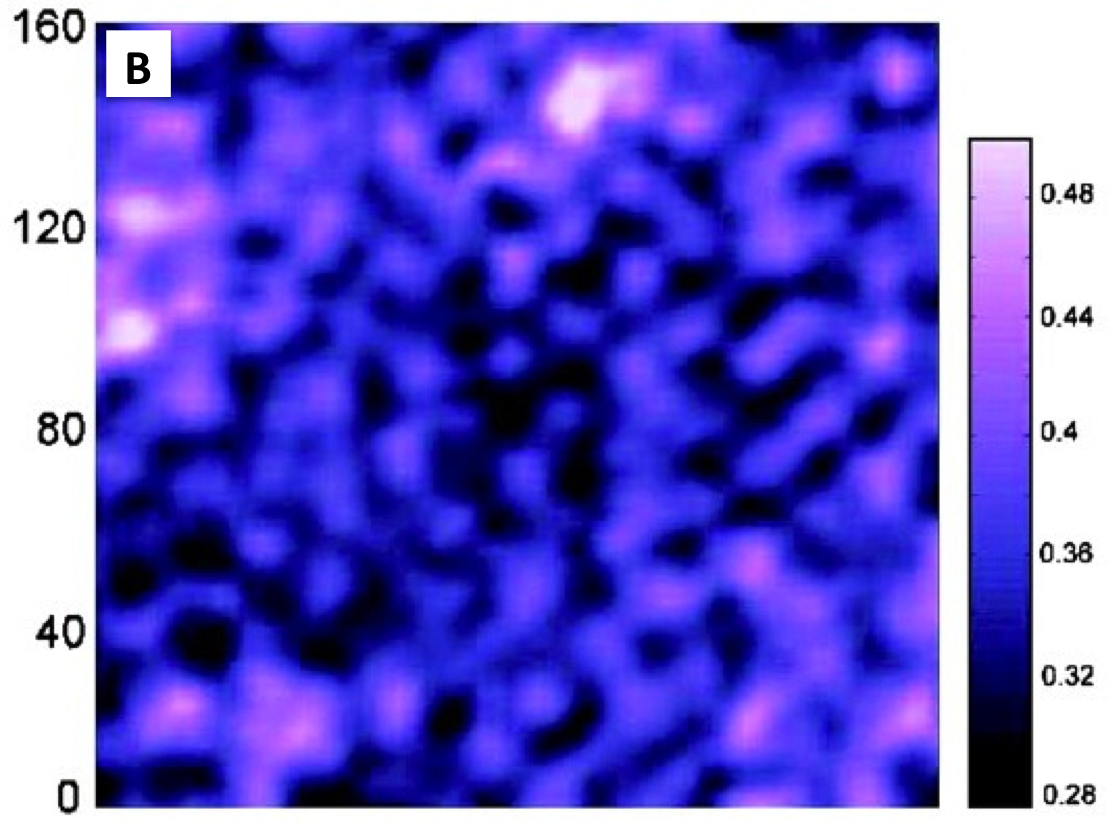}
   \includegraphics[height=.15\textheight]{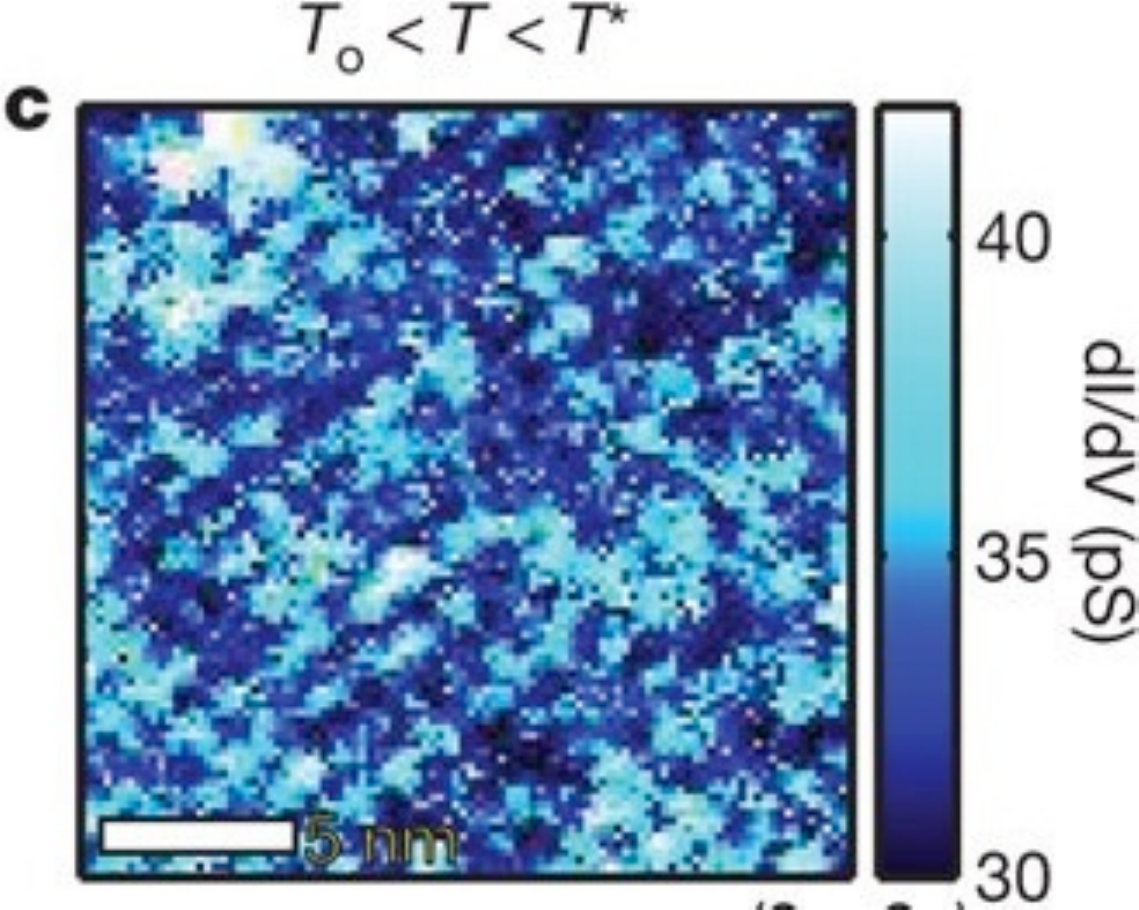}
   \includegraphics[height=.15\textheight]{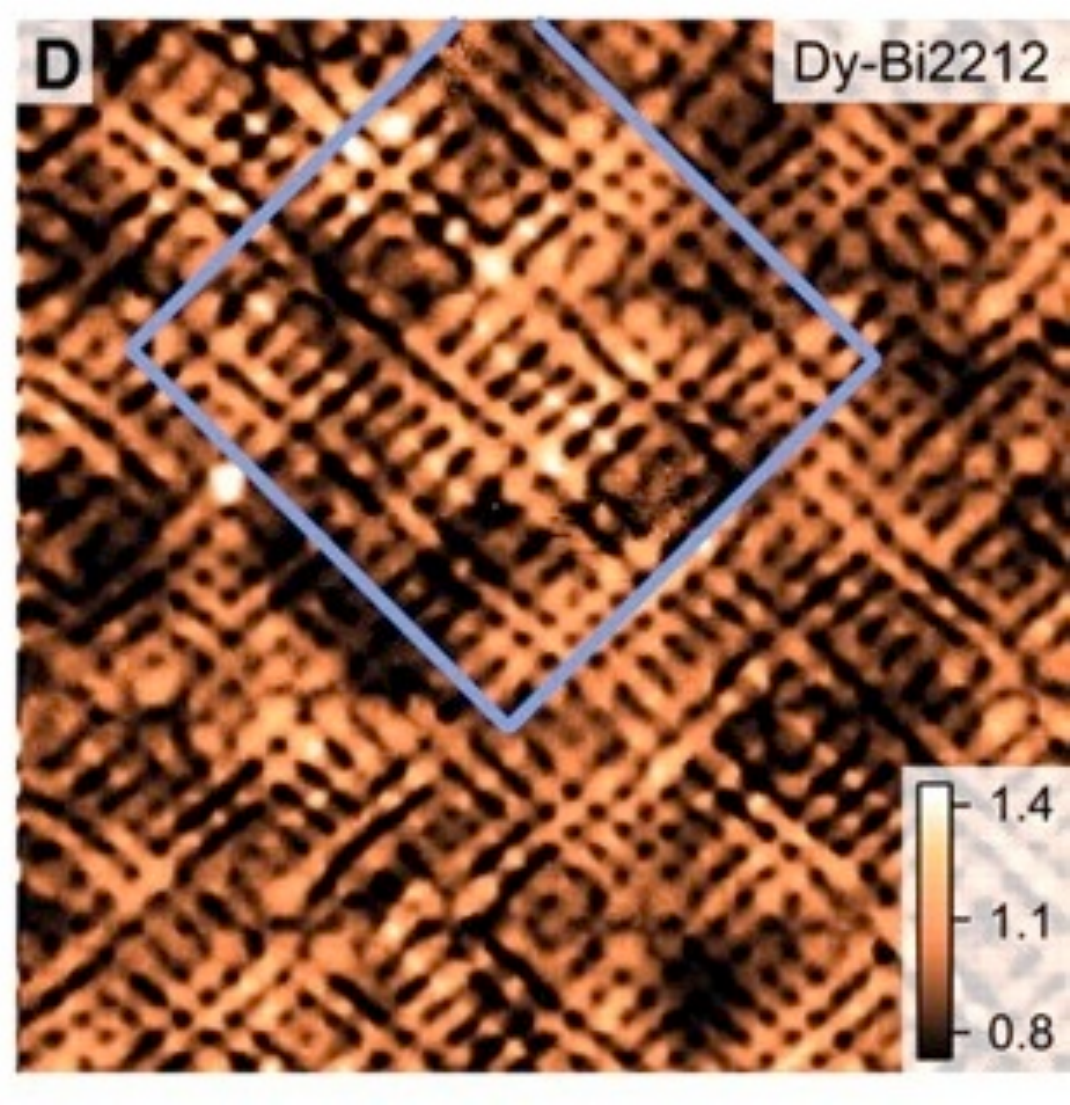}
   \caption{Patterns of locally broken spatial symmetries observed in
   atomic scale STS data from several cuprate superconductors. (a) Integrated
local density of states (LDOS) near halos of vortices in slightly over doped \BSCCO\cite{Hoffman18012002}. (b) LDOS  in underdoped \BSCCO\cite{Howald19082003} at $T< T_c$. (c) LDOS in underdoped \BSCCO for $Tc<T<T^*$\cite{parker-T*}. (d) Tunneling asymmetry in Dy-Bi2212\cite{Kohsaka09032007} that show highlighted stripe-like patterns.}
 \label{fig:pattern}
\end{figure}

In general, the challenge in establishing a broken symmetry is in devising an order parameter that can be pursued by experimental probes. Even when the target order parameter is known,  a new type of broken symmetry may require a new experimental technique. On the other hand, a new technique may call for a new order parameter that can take advantage of it. The accumulation of scanning tunneling spectroscopy (STS) data of heterogeneous patterns (see Fig.~\ref{fig:pattern}(a-b)), called for a new order parameter to turn the heterogeneous images into theoretical inputs. 

Since the concept of spontaneous symmetry breaking was established, 
bulk measurements have been perfected. At the same
time we have developed theoretical formalisms best suited to aid the connection
between bulk measurements and the physics of symmetry breaking and phase
transitions. However, apparently a heterogeneity at the nano-scale is common among strongly correlated systems. Moreover, such heterogeneity is likely due to cooperation between quenched disorder and interaction effects such as a tendency to form a ELC state. 
The existence of heterogeneity and the possibility of its intrinsic origin propelled developments in local scanning probes. 
However the lack of suitable theoretical formalism prevented the atomic scale STS data (see for instance Fig.~\ref{fig:pattern}), from providing an intermediate length scale information of ordered regions.

\begin{figure}
\centering
\subfigure[]{\includegraphics[height=.23\textwidth]{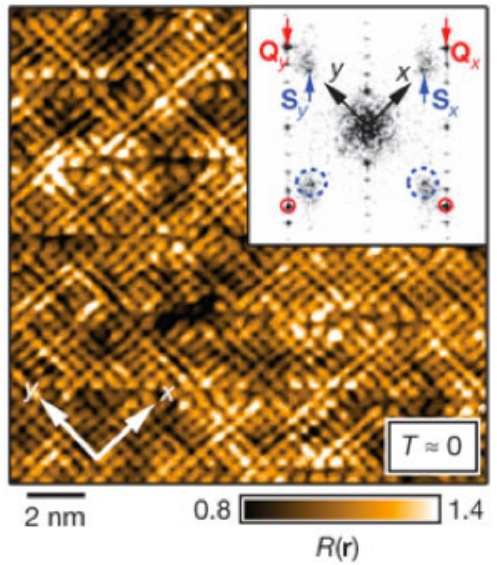} }
\subfigure[]{\includegraphics[height=.23\textwidth]{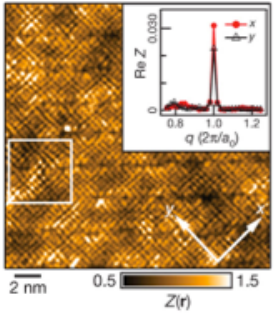} }
\subfigure[]{\includegraphics[height=.23\textwidth]{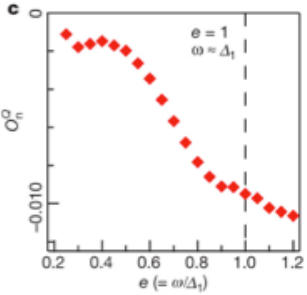} }
\subfigure[]{\includegraphics[height=.23\textwidth]{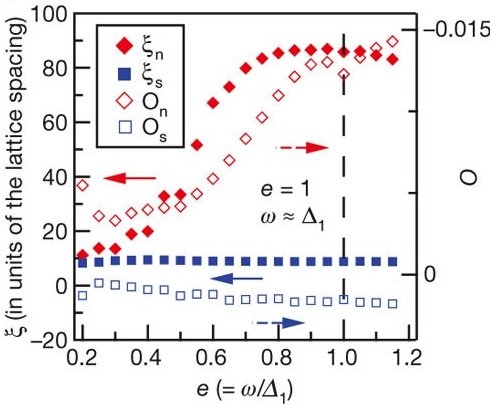}}

\caption{Evidence for nematic order in \BSCCO. (a) The real space data and its Fourier transform 
. (b) the nematic order parameter, $O_N$, from Fourier transfrom following Eq. (\ref{eq:ONe}) 
. (c) $O_N$ as a function of energy parameter $e$ as a fraction of the ``pseudogap energy scale'' $\Delta_1$.
 (d) Absence of orientational order in smectic waves (blue solid squares). 
Figures taken from ref.~\cite{Lawler:2010fk}.}
\label{fig:nematic-data}
\end{figure}

We had two goals in developing local measures of electronic liquid crystal ordering: (i) to distinguish nematic from smectic and (ii) to {\it coarse-grain} atomic scale information and define order parameter fields. We achieved both goals in Ref.~\cite{Lawler:2010fk} by going to Fourier space. In position space, patterns with a particular modulation period breaking translational symmetry of lattice and those that respect lattice translational symmetry are all superposed (see Fig. \ref{fig:pattern}). However, in Fourier space these two signals are separated as shown in the inset of Fig. \ref{fig:nematic-data}a. All information that respect lattice translation are carried by the Bragg peaks and modulation signals that break lattice translation symmetry are carried by broad peaks near $\pm\vec{S}_x\approx(\pm 3/4,0) 2\pi/a$ and  $\pm\vec{S}_y\approx(0,\pm 3/4) 2\pi/a$. By focusing on spatial variations in the STS data at each of the two atomic scale wavelengths, an IUC nematic order parameter fields and smectic order parameter fields can be defined. 

An IUC nematic order parameter associated with a real space data $M(\vec{r})$  is 
\begin{equation}
O_N[M]=\frac{1}{2}\left[\tilde{M}(\vec{Q}_y)-\tilde{M}(\vec{Q}_x)+ \tilde{M}(-\vec{Q}_y)-\tilde{M}(-\vec{Q}_x) \right],
\label{eq:ON} 
\end{equation}
where $\tilde{M}(\vec{q})$ is complex valued two-dimensional Fourier transform of $M(\vec{r})$: 
\begin{equation}
\tilde{M}(\vec{q})=\frac{1}{\sqrt{N}}\sum_{\vec{R}+\vec{d}} M(\vec{R},\vec{d})e^{-i\vec{q}_x\cdot\vec{d}}
\label{eq:FT}
\end{equation}
and $\vec{Q}_x=(2\pi/a_0,0)$ and $\vec{Q}_y=(0,2\pi/a_0)$ are two inequivalent Bragg peaks associated with unit cell dimension $a_0$. In Eq.~(\ref{eq:FT}) $\vec{R}$ is a Bravis lattice vector and $\vec{d}$ is a basis vector pointing to atomic positions within the unit cell. As defined, $O_N$ is only sensitive to signals that respects lattice periodicity and it detects an inequivalence between the $x$- and $y$-directions. Hence $O_N$ has all the features one expects of electronic nematic order parameter, which is supposed to be a headless vector\cite{nematic-review}. Furthermore, that $O_N$ is a real number is consistent with the expectation that reduction of $C_{4v}$ down to $C_{2v}$ should be through an Ising-like order parameter\cite{PhysRevB.77.184514}. However, the proposed IUC nematic order parameter Eq.~(\ref{eq:ON}) comes with two requirements on $M(\vec{r})$: (i) accurate registry of atomic sites for the phase of Fourier transform, and (ii) subatomic resolution. The latter condition is tied to the fact that $O_N\neq 0$ measures intra-unit-cell variations in $M(\vec{r})$. 

To gain insight into $O_N$ as a measure of ``intra-unit-cell'' ordering, consider a simple distribution of $M(\vec{r})$ such that $M(\vec{r})$ is non-zero only at $Cu$ sites and $O_x$, $O_y$ sites on a $CuO_2$ plane. Then 
\begin{eqnarray}
&\tilde{M}(\vec{Q}_x)&=\bar{M}_{Cu} -\bar{M}_{O_x} + \bar{M}_{O_y}\nonumber\\
&\tilde{M}(\vec{Q}_y)&=\bar{M}_{Cu} +\bar{M}_{O_x} - \bar{M}_{O_y}\nonumber\\
&O_N[M]&=\bar{M}_{O_x} - \bar{M}_{O_y} 
\end{eqnarray}
hence detection of $O_N[M]\neq0$ implies imbalance in the electronic structure at the $O_x$ sites and $O_y$ sites within each unit cell, on average. Moreover, IUC nematic order in the pseudogap phase would highlight importance of oxygen sites\cite{PhysRevB.84.144502}.  

So far we have targeted underdoped \BSCCO with this analysis scheme.  We analyzed STS data $Z(\vec{r},\omega)$ defined as\cite{Kohsaka09032007}
\begin{equation}
Z(\vec{r},\omega)\equiv \frac{g(\vec{r},\omega)}{g(\vec{r},-\omega)}=\frac{N(\vec{r},\omega)}{N(\vec{r},\omega)}
\end{equation}
as a function of reduced energy $e\equiv \omega/\Delta_1(\vec{r})$\footnote{The reduced energy scale helps one to consider energy scales relative to the strongly position dependent pseudogap scale $\Delta_1(\vec{r})$. See Ref\cite{Lawler:2010fk} and references therein.} for various underdoped samples. We confirmed that 
the magnitude of $O_N(e)$
\begin{equation}
O_N(e)\equiv \frac{Re \tilde{Z}(Q_y,e)-Re \tilde{Z}(Q_x,e)}{\bar{Z}(e)},
\label{eq:ONe}
\end{equation}
where $\bar{Z}(e)$ is the spatial average of $Z(\vec{r},e)$,  grows with the reduced energy until $e\approx 1$ reaches the pseudogap scale. Fig.~\ref{fig:nematic-data}(c) shows one such example. 

Lets us now turn to the smectic modulations. A smectic order parameter can be defined in analogy to Eq.~(\ref{eq:ONe}) as
\begin{equation}
O_S(e)\equiv \frac{Re \tilde{Z}(S_y,e)-Re \tilde{Z}(S_x,e)}{\bar{Z}(e)}.
\label{eq:OSe}
\end{equation}
We note that $O_S$ above focuses on whether modulation in one direction is dominant over modulation in another direction. Each modulation component has an amplitude and a phase, hence they should each be associated with a complex order parameter. We will revisit this issue in the next section. However, the simplified view of translational symmetry breaking features through Eq.~(\ref{eq:OSe}) already led to a surprising observation. Fig.~\ref{fig:nematic-data} (d) shows a clear contrast between $O_N(e)$ and $O_S(e)$: while $O_N(e)$ becomes robust at $e\approx1$, $O_S(e)$ remains small through out. This is particularly interesting since strong local modulation is what stands out the most in Fig.~\ref{fig:nematic-data}(a) to bare eyes. 

\begin{figure}
\centering
\subfigure[]{\includegraphics[height=.4\textwidth]{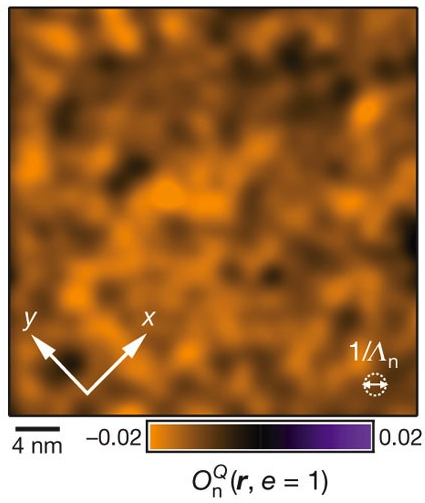}} $\qquad\qquad$
\subfigure[]{\includegraphics[height=.4\textwidth]{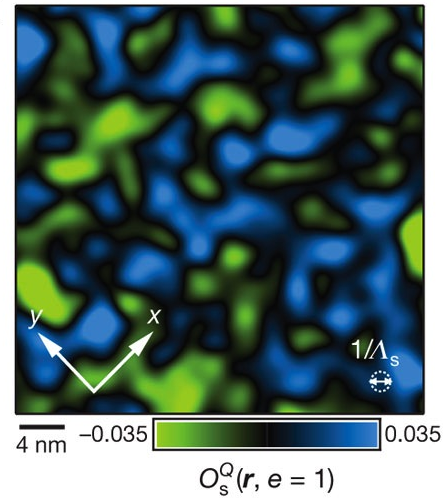}} 
\caption{Long wave length order parameters extracted from Fig. \ref{fig:nematic-data}a. (a) $O_N(\vec{r}, e=1)$ showing a uniform bias towards the negative value of -0.02 and (b) $O_S(\vec{r}, e=1)$ showing strong spatial fluctuations between two extreme values. }
\label{fig:fields}
\end{figure}

In order to resolve the mismatch between what stands out to our eyes and what $O_N(e)$ and $O_S(e)$ shows, it is important to capture spatial fluctuations in the IUC nematic and smectic order parameters ($O_N(e)$ and $O_S(e)$ are image-wide averages).  To this end, Fourier filtering technique are very useful. With the definition of global orders Eqs.~(\ref{eq:ONe}-\ref{eq:OSe}), we can define order parameter fields $O_N(\vec{r}, e)$ and $O_S(\vec{r},e)$ whose average yields $O_N(e)$ and $O_S(e)$ respectively. A coarse grained field $\tilde{Z}(\vec{Q};\vec{r})_\Lambda$ for a Fourier peak centered at $\vec{Q}$ can be obtained by ``cutting out'' the peak and shifting the center of the peak to the origin in Fourier space. This second step removes sub-atomic scale variation.  Finally upon inverse Fourier transform a coarse grained field configuration is obtained. One practical issue in using this procedure was the slow piezo drift in data. We developed a scheme to correct for such drift when simultaneous topograph information is available by introducing a displacement field $\vec{u}(\vec{r})$ (see appendix). With this $\tilde{Z}(\vec{Q};\vec{r})_\Lambda$ becomes
\begin{eqnarray}
\tilde{Z}(\vec{Q},\vec{r})_\Lambda&\equiv& \sum_{\vec{r}'} Z(\vec{r}')e^{ i\vec{Q}\cdot(\vec{r'}-\vec{u}(\vec{r'}) )} f_\Lambda(\vec{r}'-\vec{r}) \\
&\approx& \frac{1}{\sqrt{N}}\sum_{\vec{k}} \tilde{Z} (\vec{Q}-\vec{k})e^{i \vec{k}\cdot (\vec{r}-\vec{u}(\vec{r}) )} e^{-k^2/2\Lambda^2}  
\end{eqnarray}
where $f_\Lambda(\vec{r})\equiv\frac{\Lambda^2}{2\pi} e^{ -\Lambda^2|\vec{r}|^2/2}$ is used to implement the cutoff at length scale $1/\Lambda$. For  $O_N(\vec{r}, e)$ and $O_S(\vec{r}, e)$ we set the cutoff to the $3\sigma$ radius of the Bragg peaks and $\vec{S}_x$, $\vec{S}_y$ peaks respectively. The resulting maps of 
$O_N(\vec{r}, e=1)$ and $O_S(\vec{r}, e=1)$ of a representative under-doped sample are shown in Fig~\ref{fig:fields}. These maps confirm that severe spatial fluctuation in $O_S(\vec{r}, e)$ at all $e$ suppresses $O_S(e)$ though locations with large local $|O_S(\vec{r}, e)|$ grab our attention. On the other hand for large enough $e$, $O_N(\vec{r},e)$ only fluctuates mildly around a finite global average $O_N(e)$.

\section{Coupling between Nematic and Smectic}
\label{sec:coupling}
It has been argued based on symmetry grounds and analogies with liquid crystals that one mechanism of the formation of an electronic nematic phase is through disordering the smectic modulations\cite{Kivelson:1998nature}. This mechanism was the starting point for a phenomenological model of fluctuating stripe phenomena in YBCO~\cite{PhysRevLett.104.106405}. However, there is no unambiguous realization of this mechanism yet. Further, there is no theory of the melting of an electronic smectic phase via quantum fluctuations. However, by direct observation of the coexistence of long-range nematic order and the disordered smectic modulations reported in Ref.~\cite{Lawler:2010fk}, such a melted smectic phase seems the most reasonable description of \BSCCO. We therefore have concrete testing bed for a coupling between nematic and smectic orders: the subject of Ref.~\cite{Mesaros22072011} and this section.

In the above context, two questions are: (i) how do severely disordered smectic modulations and long range nematic order coexist and (ii) what do we gain by being able to map the spatially fluctuating order parameters.  In order to answer these questions we constructed a field theory describing a nematic field and modulation fields based on symmetry principles: a Ginzburg-Landau free energy. We then tested whether this field theory captures the essence of the nematic-smectic coexistence reported in Ref.~\cite{Lawler:2010fk}. For this, a scalar field
\begin{equation}
\delta O_N(\vec{r})\equiv O_N(\vec{r},e=1)-O_N(e=1)
\label{eq:delON}
\end{equation}
represented local nematic fluctuation away from the global expectation value. On the other hand, two complex fields $\psi_1(\vec{r})$ and $\psi_2(\vec{r})$ represented disordered modulations with wave vectors $\vec{S}_x$ and $\vec{S}_y$ each\cite{Mesaros22072011}. 

Let us start with the GL free energy for the modulations:
\begin{eqnarray}
  \label{eq:0}
  F_{S}[\psi_1,\psi_2] &=&\int\textrm{d}\vec{r}\Bigl[{a}_{x,1}|\dx\psi_1|^2+{a}_{y,1}|\dy\psi_1|^2+m_1|\psi_1|^2\\\nonumber &\quad&\quad\quad\;+{a}_{x,2}|\dx\psi_2|^2+{a}_{y,2}|\dy\psi_2|^2+m_2|\psi_2|^2\Bigr], 
\end{eqnarray}
where we kept terms up to quadratic order in $\psi_i$. $m_1$, $m_2$ are positive as we found
$\psi_i(\vec{r})$ to average to zero\cite{Lawler:2010fk}. In Eq.~(\ref{eq:0}) we assumed 
an orthorombic crystal symmetry ($C_{2v}$), based on the observation of
$\fz\neq 0$.

The GL free energy $F_n[\df]$ for the nematic fluctuation $\df$ is that for an Ising field fluctuation in an ordered state:
\begin{equation}
  \label{eq:2}
F_{N}[\df]=\int\textrm{d}\vec{r}^2\left[\sum\limits_{i=x,y}(\di\df)^2+\frac{1}{\xi^2_N}\df^2\right],
\end{equation}
where $\xi_N$ is the nematic fluctuation correlation length and we assumed isotropic nematic fluctuation for simplicity. 

Now we turn to the coupling. Since the modulation fields $\psi_1$ and $\psi_2$ are complex, the nematic fluctuation scalar field can couple either to the amplitude or phase of the modulation fields. The lowest order amplitude coupling terms are
\begin{equation}
  \label{eq:7}
  \dcoup_s\df(\vec{r})|\psi_s(\vec{r})|^2,
\end{equation}
with coupling constants $\dcoup_s$ for $\psi_s$, $s=1,2$. These terms represent local enhancements of smectic amplitude fluctuations caused by the nematic fluctuation. The phase coupling comes from 
nematic fluctuation inducing a local stretching or compression of the smectic modulation:
\begin{equation}
  \label{eq:5}
  \vec{S}_s\rightarrow\vec{S}_s+\vec{c}\df(\vec{r}),
\end{equation}
where the vector $\vec{c}$ is a phenomenological coupling constant. In terms of the modulation fields $\psi_s(\vec{r})$, the shift of the modulation wave vector Eq.~\ref{eq:5} amounts to a covariant derivative:
\begin{equation}
  \label{eq:6}
  \di\psi_s(\vec{r})\rightarrow\left(\di+i c_i\df(\vec{r})\right)\psi_s(\vec{r}).
\end{equation}

The final form of the GL functional containing all the symmetry allowed lowest order (up to quadratic in each field)  terms  is
\begin{eqnarray}
  \label{eq:03}
  &&F_{GL}[\df,\psi_1,\psi_2]=F_N[\df]+\int\textrm{d}\vec{r}^2 \sum\limits_{ s=1,2}\Bigl[{a}_{x,s}|(\dx+i c_x\df)\psi_s|^2\nonumber\\
&&\quad \qquad\qquad+{a}_{y,s}|(\dy+i c_y\df)\psi_s|^2+{m}_s|\psi_s|^2+\dcoup_s\df|\psi_s|^2\Bigr]\nonumber\\  
  &&\quad=
  \int\textrm{d}\vec{r}^2 \sum\limits_{ s=1,2}\sum\limits_{i=x,y}\left[\bcoup_{i,s}\df|\psi_s|^2\di\varphi_s+\dcoup_s\df|\psi_s|^2+\gcoup_s\df^2|\psi_s|^2\right]\nonumber\\
&&\quad \qquad+F_N[\df]+F_{S}[\psi_1,\psi_2],
\end{eqnarray}
where we introduced compact labels for the coupling constants $\bcoup_{i,s}=\sum_{i=x,y}a_{i,s}c_i$ and $\gcoup_{i,s}=\sum_{i=x,y}a_{i,s}c^2_i$. 
{\color{blue}*}We can make two observations. First, the coupling between the modulation phase and fluctuations in the nematic order Eq.~(\ref{eq:6}) resembles the coupling between phase of smectic waves and the nematic director in liquid crystals\cite{deGennes1972753,Toner1981}, except that the nematic director is a headless vector which can lie along any direction while $\delta O_N$ is a scalar needing an auxilliary vector $\vec{c}$. Second, as the phase coupling is a gradient coupling, one would expect the amplitude coupling to dominate in the homogeneous limit, in which case the strong fluctuation in the smectic modulations would likely disorder the nematic phase. However, pseudogap states in \BSCCO  outwits such guess due to inhomogeneity and the dominance of the phase coupling saves the long range nematic order. 

\begin{figure}
\includegraphics[width=0.34\textwidth]{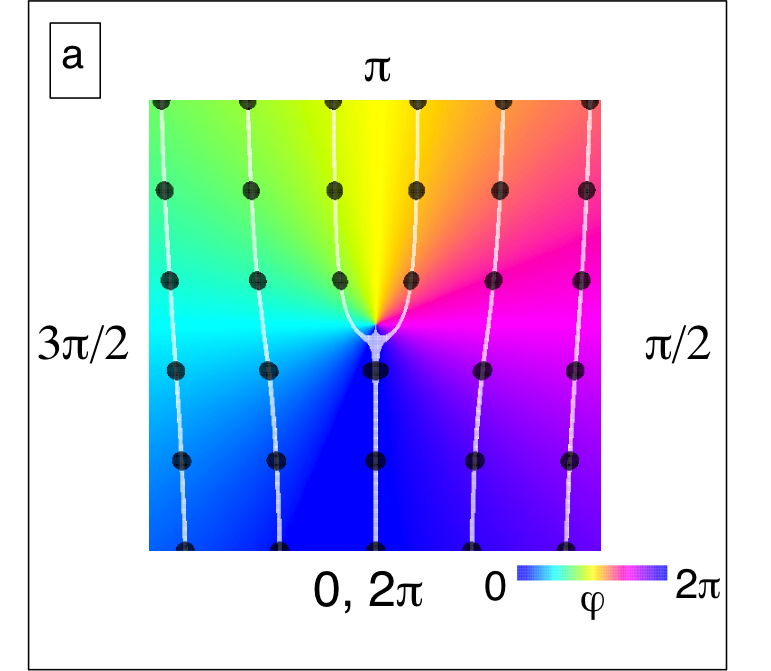}
\includegraphics[width=0.3\textwidth]{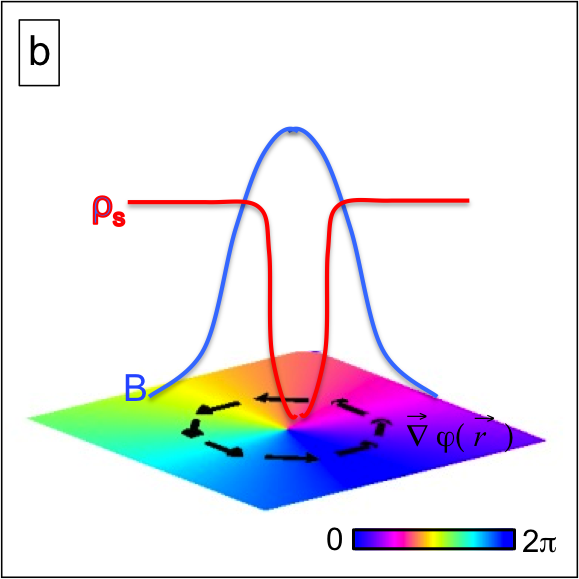}
\includegraphics[width=0.3\textwidth]{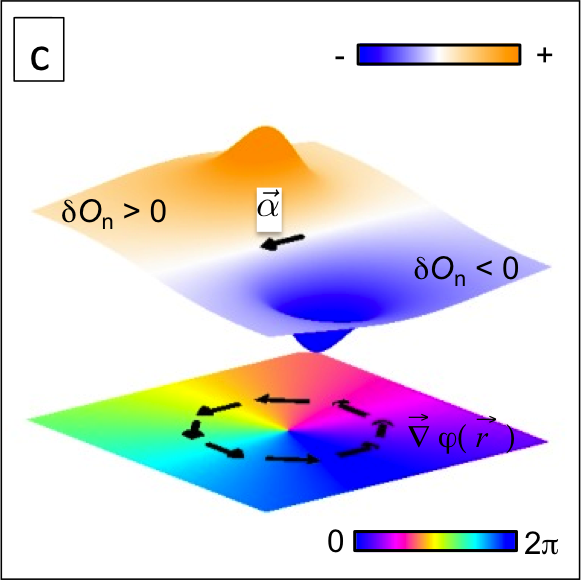}
\caption{vortex and topological defect}
\label{fig:vortex}
\end{figure}

A key new insight  the phase coupling in Eq.~(\ref{eq:03}) offers regards how $\delta O_N$ is affected by a stripe dislocation. A stripe dislocation is a topological defect in modulations $\psi_{1}(\vec{r})e^{i \vec{S_x}\cdot\vec{r}}=|\psi_{1}| e^{i[\vec{S_{x}}\cdot\vec{r}+\varphi_1(\vec{r})]}$ and  $\psi_{2}(\vec{r})e^{i \vec{S_x}\cdot\vec{r}}=|\psi_{2}| e^{i[\vec{S_{x}}\cdot\vec{r}+\varphi_2(\vec{r})]}$ associated with $\pm 2\pi$-multiple windings of the phase fields $\varphi_1$ and $\varphi_2$. At a stripe dislocation, a ridge of modulation terminates analogous to a crystal dislocation at which a line of atoms terminate(see Fig.~\ref{fig:vortex}a). 
To see the implication of the phase coupling  in Eq.~(\ref{eq:03}), 
 it is useful to note that if we were to replace $\vec{c}\delta O_N(\vec{r)}$ by $\frac{2e}{h}\vec{A}(\vec{r})$ where $\vec{A}(\vec{r})$ is the electromagnetic vector potential, Eq.~(\ref{eq:03}) becomes the GL free energy of a superconductor under magnetic field. For the latter case, minimization of the GL free energy in the long distance limit yields  $\vec{A}(\vec{r})=\frac{h}{2e}\vec{\nabla}\varphi(\vec{r})$ relative to the center for the vortex and thus quantization of its associated magnetic flux.  Analogously, minimization of the GL free energy Eq.~(\ref{eq:03}) implies $\delta O_N(\vec{r})\propto \vec{\alpha}\cdot\vec{\nabla}\varphi$ where $\varphi$ denotes the phase of a modulation field(see Fig.~\ref{fig:vortex}b). This means that $\delta O_N(\vec{r})$ will vanish along the contour in the direction of  $\vec{\alpha}$ that passes through the core of a stripe dislocation.

\begin{figure}
\centering
\includegraphics[width=0.32\textwidth]{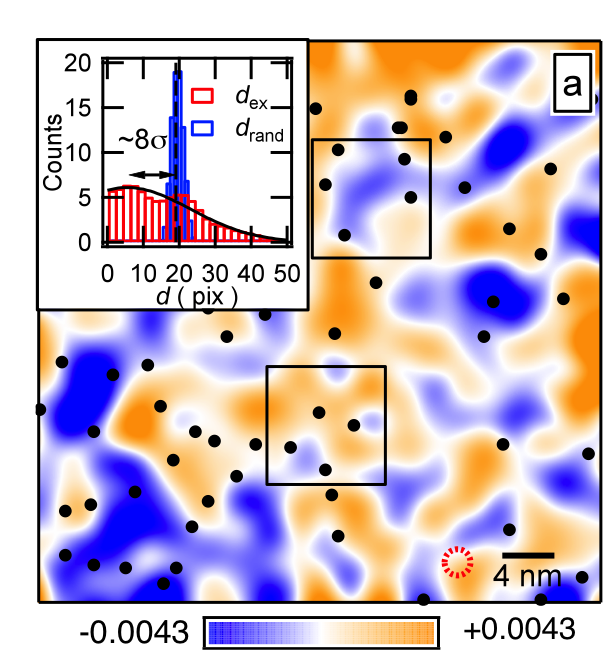}
\includegraphics[width=0.3\textwidth]{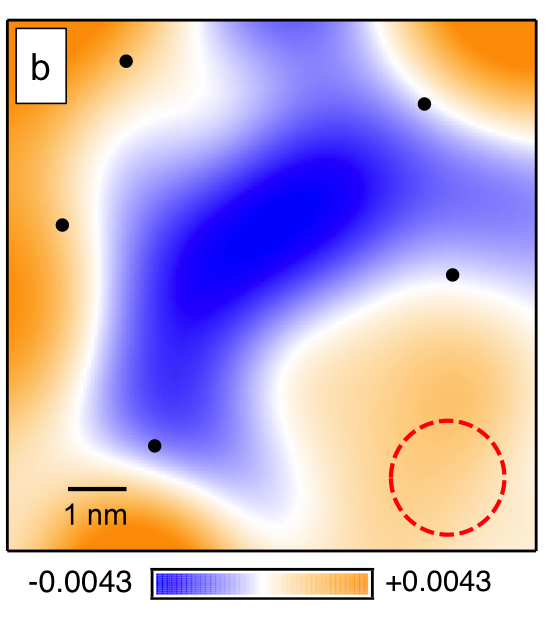}
\includegraphics[width=0.33\textwidth]{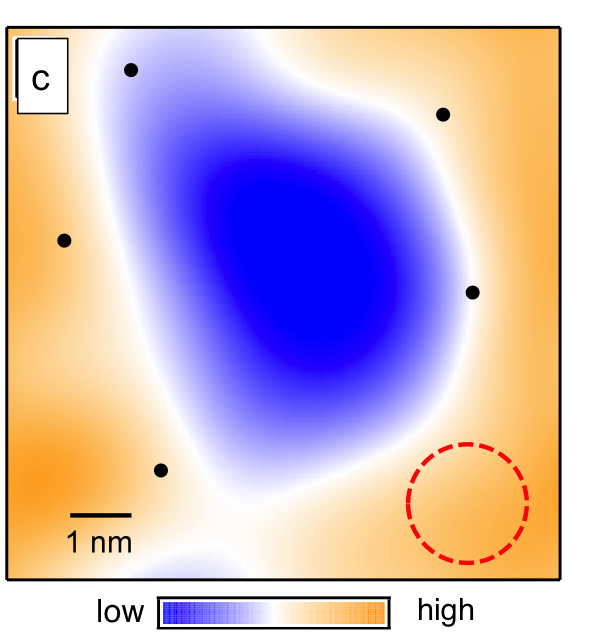}
\caption{(a) The overlay of $\delta O_N(\vec{r})$ on top of a map of stripe dislocation positions. The inset shows the distribution of distances between each topological defects and its nearest $\delta O_N(\vec{r})=0$ contour in red histogram. This is compared to the expected average distance in blue histogram. (b) Blow-up of the box to the immediate right of the inset. (c) The simulated $\delta O_N(\vec{r})$ given the positions of the stripe dislocations. Figures are  taken from Ref.~\cite{Mesaros22072011}
}
\label{fig:dist}
\end{figure}

To test the applicability of the GL functional Eq.~(\ref{eq:03}), we first mapped out all stripe dislocations and took the statistics of distance from a dislocation to the nearest point on the contour of vanishing nematic fluctuation $\delta O_N(\vec{r})=0$, i.e. the contour along which $O_N(\vec{r})=O_N$. There is a very strong tendency for the distance to the nearest $O_N(\vec{r})=O_N$ contour to be small (Fig.~\ref{fig:dist}a). Further we compared the nematic fluctuation $\delta O_N(\vec{r})$ in a region (Fig.~\ref{fig:dist}b) to a ``simulated'' $\delta O_N(\vec{r})$ based on the positions of the stripe dislocations and the GL functional Eq.~(\ref{eq:03}) (Fig.~\ref{fig:dist}c). For the latter, we modeled each stripe dislocation using  
\begin{equation}
  \label{eq:3}
|\psi_s(\vec{r})|^2\vec{\nabla} \varphi_s(\vec{r})=\left(1-\exp{(-|\vec{r}|^2/\xi_S^2)}\right)\vec{\nabla}\theta(\vec{r}),
\end{equation}
where $\theta$ is the polar angle in the plane when the origin is set to be the center of the dislocation and $\xi_S$ the coherence length of the modulation field. Then we use this as a source that generates  $\delta O_N(\vec{r})$ configuration according to its coupling to the modulation fields in Eq.~(\ref{eq:03}):
\begin{equation}
  \label{eq:4}
\dfd(\vec{r})=\int\textrm{d}^2\vec{r}' G(\vec{r}-\vec{r}') \sum\limits_{
  s=1,2} |\psi_s(\vec{r}')|^2\left\{\sum\limits_{i=x,y}\bcoup_{i,s}\di \varphi_s(\vec{r}')+\dcoup_s\right\}
\end{equation}
where $G(\vec{r})=K_0(|\vec{r}|/\xi_N)$, with $K(r)$ the Bessel function of the second kind from $F_N[\delta O_N]$ in Eq.~(\ref{eq:2}). Then we superposed contributions sourced by each stripe dislocations and determined the coupling constants $ (\bcoup_{x,1},\bcoup_{x,2},\bcoup_{y,1},\bcoup_{y,2}, \beta_1, \beta_2)=(4,16,4,-4,8,2)$ with $\xi_S/\xi_N=0.1$ \footnote{Note that we replaced the modulation fields by a superposition of modulation field profile near isolated dislocations,  whose locations and windings are taken from experimental data. We checked that the resulting smectic fields are almost identical to the experimental ones, meaning that it is a reasonable assumption that the dislocations determine the fluctuation of smectic fields. }. While this is not a full solution to the coupled GL functional Eq.~(\ref{eq:03}), rather a mean-field approximation, good comparison between Fig.~\ref{fig:dist}b and c (cross-correlation coefficients of 62\%) demonstrates the validity of the GL functional Eq.~(\ref{eq:03}). 

We close this section with two remarks. (1) We first note that we only skimmed the surface of the rich physics of a disordered smectic phase. We did not address the question of whether the defects are intrinsically or extrinsically generated. That is, we have not addressed the origin of the formation of these topological defects. Correlations with dopant disorder could shed some light on this question. (2) Nevertheless, we settled the question of how it is possible for a disordered stripe field to coexist with long range IUC nematic in underdoped \BSCCO. Further our analysis offers a starting point for a microscopic theory of quantum melting of stripes.

\section{IUC Orders in the Emery Model}
\label{sec:emery}
Various intra-unit-cell symmetry breaking possibilities are among accumulating experimental evidence for symmetry breaking in underdoped cuprates.
Neutron scattering experiments discovered a subtle
staggered magnetic order in the pseudo-gap region of
YBCO\cite{fauque:2006} and Hg-compounds\cite{li:2008}
that could be accounted for by either so-called nematic-spin-nematic
order\cite{oganesyan:2001, wu:2007} or circulating current
loops\cite{varma:2006b}. On the other hand, neutron
scattering~\cite{hinkov:2008} and Nernst effect~\cite{daou:2010}
measurements on YBCO as well as SI-STM on BSCCO~\cite{Lawler:2010fk}
point towards an electronic nematic state. 
All these states retain the
translational symmetry of the underlying crystal and can thus
naturally be described by breaking intra-unit-cell (IUC) symmetries. 
However, theories of IUC ordering mostly focus on 
one particular ordering within a simplified
model, each aimed at an ordering of interest.
Nematic and nematic-spin-nematic order have only been studied in one-band models \cite{yamase:2000c,oganesyan:2001, wu:2007, kee:2003,metzner:2003, yamase:2005,
halboth:2000, gull:2009,okamoto:2010} or in the extreme limit of infinite interactions\cite{kivelson:2004}. Loop currents, being more dependent on an IUC picture, have been studied in a mean-field picture with additional assumptions\cite{varma:2006b} or numerically on small clusters or ladders\cite{chudzinski:2007, greiter:2008, weber:2009}.

In this section we review a comprehensive investigation of IUC-ordering
possibilities following Ref.~\cite{PhysRevB.84.144502}.
\footnote{Other IUC-ordering possibilities were considered e.g. by Sun et al.~\cite{sun:2008}}
The observation of the importance of oxygen sites in detecting IUC nematic in Ref.~\cite{Lawler:2010fk} motivated this study of a microscopic mechanism for IUC nematic order in the so-called Emery
model\cite{emery:1987}. We summarize the results of this self-consistent mean-field theory on three distinct IUC orders: 
nematic, nematic-spin-nematic,
 and loop currents~\ref{tab:sym}. 
\begin{table}[h]
  \centering
  \begin{tabular}{c|c|c|c|c|c}
    & $C_4$ & $\mathcal{I}$ & $\mathcal{T}$ & $C_4\circ \mathcal{T}$ & $\mathcal{I}\circ\mathcal{T}$\\
    \hline
    \hline
    nematic &$\quad\times\quad$ & $\quad\phantom{-}\quad$ & $\quad\phantom{-}\quad$ & $\quad\times\quad$ &$\quad\phantom{-}\quad$\\
    \hline
    nematic-spin-nematic &$\times$  &  & $\times$ & $ $ & $ $\\
    \hline
    $\Theta_{II}$ loop current & $\times$ & $\times$ & $\times$ & $\times$ &
  \end{tabular}
  \caption{The broken symmetries distinguishing the different IUC orderings with $\times$ denoting symmetries broken in the respective phase. For simplicity, we restrict the table to the fourfold rotation $C_4$, the inversion $\mathcal{I}$, time-reversal operation $\mathcal{T}$ as well as combinations thereof. For simplicity, only fourfold rotations, inversion, time reversal and combinations of these are shown. }
  \label{tab:sym}
\end{table}

\begin{figure}
\centering
\includegraphics[width=.35\textwidth]{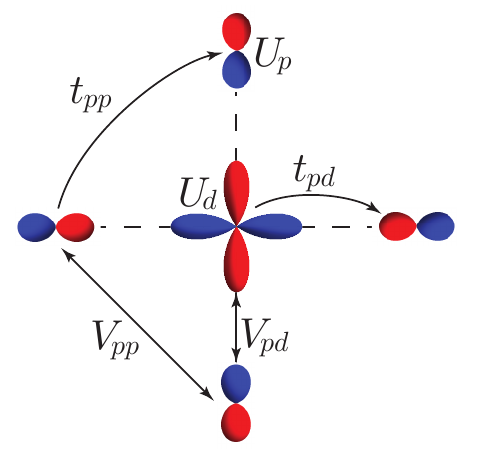}
 \caption{The unit cell of the CuO$_2$ plane with the copper $d_{x^2-y^2}$ in the middle surrounded by the oxygen $p_x$ and $p_y$ orbitals. Also shown are the different hopping as well as interaction parameters used in the Emery model.}
\label{fig:unitcell}
\end{figure}

The kinetic part of the Emery model~\cite{emery:1987} describing hopping
of holes in the CuO$_2$ plane is
\begin{eqnarray}
  &&\mathcal{H}_{0} = -t_{pd}\sum_{i,s}\sum_\nu(\hat{d}^{\dag}_{i,s}\hat{p}^{\phantom{\dag}}_{ i+\hat{\nu}/2, s} + {\rm h.c.})
  - t_{pp}\sum_{i, s}\sum_{\langle\nu,\nu'\rangle}(\hat{p}^{\dag}_{ i+\hat{\nu}/2, s}\hat{p}^{\phantom{\dag}}_{ i+\hat{\nu}'/2, s} + {\rm h.c.})\nonumber\\ 
 &&\quad\quad - \mu \sum_{i,s}\hat{n}^{d}_{i,s} - \frac12(\mu - \Delta)\sum_{i,s}\sum_{\nu}\hat{n}^{p}_{i+\hat{\nu}/2,s}
  \label{eq:hopping}
\end{eqnarray}
with $t_{pd}$ and $t_{pp}$ the Cu-O and O-O hopping integrals. Here, $\hat{d}^{\dag}_{i,s}$ creates a hole
in the copper $d_{x^2-y^2}$ orbital at site $i$ with spin $s$,
$\hat{p}^{\dag}_{ i+\hat{\nu}/2, s}$ creates a hole in the oxygen $p_\nu$ orbital
at the site $i+\hat{\nu}/2$
for $\nu = x, y$, and
$\hat{n}^{d}_{i,s}$,
$\hat{n}^{p}_{ i+\hat{\nu}, s}$
are the corresponding number operators.
The Cu sites $i$ form a square
lattice with unit vectors $\hat{x}$ and $\hat{y}$, and the total number of lattice sites is $N$. The chemical potential $\mu$
and the charge transfer energy $\Delta$ control the total and relative
Cu/O hole densities, and $\langle\nu, \nu'\rangle$ point to neighboring oxygen sites.
We consider an interaction Hamiltonian including on-site interactions with strengths $U_{d}$ and $U_{p}$ as well as nn interactions, $V_{pd}$ and $V_{pp}$,
\begin{eqnarray}
 && \mathcal{H}' = U_{d}\sum_{i}\hat{n}_{i\uparrow}^{d}\hat{n}_{i\downarrow}^{d} + \frac{U_{p}}{2}\sum_{i, \nu}\hat{n}_{ i+\hat{\nu}/2,\uparrow}^{p}\hat{n}_{ i+\hat{\nu}/2,\downarrow}^{p}
  + V_{pd}\sum_{i, \nu}\sum_{s, s'}\hat{n}_{i,s}^{d}\hat{n}_{ i+\hat{\nu}/2, s'}^{p}\nonumber\\
  &&+ V_{pp}\sum_{i}\sum_{\langle\nu,\nu'\rangle}\sum_{s,s'}\hat{n}_{ i+\hat{\nu}/2, s}^{p}\hat{n}_{i+\hat{\nu}/2',s'}^{p}.
  \label{eq:interaction}
\end{eqnarray}
The different orbitals and parameters of the model are shown in
Fig.~\ref{fig:unitcell}. Setting $t_{pd}=1$, we fix the energy scale
in the following.

For a self-consistent mean-field phase diagram, the interaction terms Eq.~\ref{eq:interaction} should be decomposed. For nematic and nematic-spin-nematic, only the Hartree channel decomposition is necessary, using the {\it IUC nematic} order parameter
\begin{equation}
  \eta\equiv (n_{x\uparrow}^p+n_{x\downarrow}^{p}) -(n_{y\uparrow}^p + n_{y\downarrow}^p),
  \label{eq:eta}
\end{equation}
and the {\it nematic-spin-nematic} order parameter
\begin{equation}
  \eta_s\equiv (n_{x\uparrow}^p-n_{x\downarrow}^{p}) -(n_{y\uparrow}^p - n_{y\downarrow}^p).
  \label{eq:etas}
\end{equation}
Here $n_{x,s}^p$ and $n_{y,s}^p$ each refers to spin $s-$hole occupation in the oxygen $p_x$ orbital and $p_y$ orbital respectively. On the other hand, the loop-current order requires Fock decomposition of $V_{pd}$ and $V_{pp}$ interactions. 

The analysis of Ref.~\cite{PhysRevB.84.144502} made two key contributions that can serve as stepping stones for going beyond mean-field theory. First is the mean-field phase diagrams for each IUC ordering possibilities, which can guide the exploration of the large parameter space. Second is the qualitative understanding of the role of each interaction terms in promoting or suppressing certain IUC order. 

\begin{figure}[tb]
\centering
    \subfigure[]{\includegraphics[width=.51\textwidth]{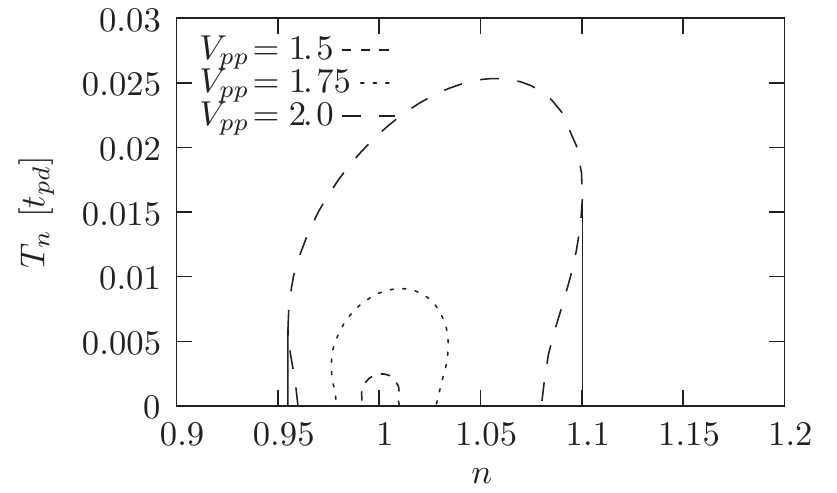}}
   \subfigure[]{ \includegraphics[width=.47\textwidth]{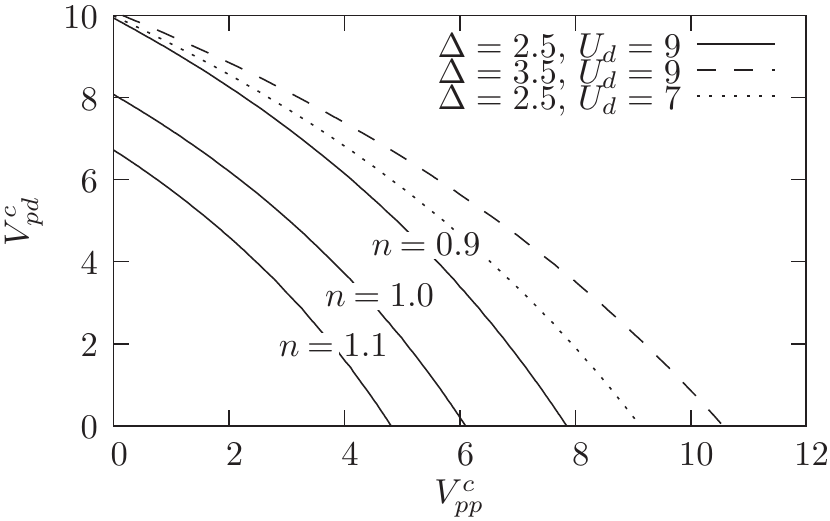}}
  \caption{(a)Doping dependence of the nematic transition temperature $T_n$,   for the different values of the O-O
    nearest-neighbor interaction $V_{pp} = 2$, $1.75$ and $1.5$. At
    low temperature, there would be first-order transitions, only
    shown for $V_{pp} = 2$ by the solid lines, before the normal state
    becomes unstable (dashed lines). The effect of
increasing $U_d$ is almost the same as increasing $V_{pp}$. (b)Critical interactions $(V_{pp}^c, V_{pd}^c)$ for $U_d = 9$, $U_p = 3$, $t_{pp} = 0.1$, $\Delta = 2.5$ and different hole densities. The dashed and dotted lines for $n=0.9$ illustrate the influence of the Cu on-site interaction and the charge transfer gap.}
  \label{fig:Upc}
\end{figure}
We first discuss the mean-field phase diagram. 
For nematic and nematic-spin-nematic, the critical interaction strength has a non-monotonic doping dependence due to the van Hove singularity. The resulting phase diagram has a dome shape with a maximum $T_c$. On the other hand, the critical interaction strengths here are
monotonically decreasing with increasing hole density (see
Fig.~\ref{fig:Upc}(b)). This is due to the fact that the current
loop in a mean-field approach arises due to a Fock-type 
rather than Hartree-type decoupling and hence not 
a Stoner-type instability.
As the whole dispersion is altered by the decoupling, increasing the hole density in the lowest band increases the tendency towards loop currents. In order to find a phase diagram as found in the cuprates, additional assumptions to the model have to be made, such as a density-dependent hopping, e.g. of the form $t_{pd}\rightarrow t_{pd}|x|$ with $x = n-1$, as in Varma's analysis.~\cite{varma:2006b}

Now we turn to the role of various interaction terms in promoting or suppressing IUC orders. Different interaction parameters affect the various instabilities differently: while the O on-site repulsion $U_{p}$ only favors the nematic-spin-nematic phase and the Cu-O repulsion $V_{pd}$ the loop currents, the nearest-neighbor O-O repulsion $V_{pp}$ helps both, the nematic and the loop-current phase (see Tab.~\ref{tab:params} for a summary of all the model parameters). However, the Cu on-site interaction $U_{d}$ promotes all the studied orderings by shifting more holes to the oxygens. The charge transfer gap $\Delta$ has the opposite effect.
\begin{table}[ht]
  \centering
  \begin{tabular}{c|c|c|c|c|c|c}
    & $U_d$ & $U_p$ & $V_{pd}$ & $V_{pp}$ & $t_{pp}$ & $\Delta$\\
    \hline
    \hline
    nematic & + & - & - & + & - & -\\
    \hline
    nematic-spin-nematic & + & + & - & - & - & -\\
    \hline
    $\Theta_{II}$ loop current & + & -& + & + & - & -
  \end{tabular}
  \caption{Summary of the effect of the different parameters in the Emery model on the different IUC orders, where + denotes a parameter that helps a specific order and a - denotes a hindering parameter.}
  \label{tab:params}
\end{table}

The critical interaction strength within the above mean-field theory are unrealistically large. However, to actually assess whether IUC ordering can occur in realistic setting is beyond the applicability of a mean-field theory. Rather the results summarized above should serve as a starting point for a more sophisticated calculation, for instance, extension of the calculations in~\cite{halboth:2000, okamoto:2010} to the case of three bands. Nevertheless, we can draw some conclusions about the competition or coexistence of the IUC-ordered phases from lessons within mean-field theory. One interesting lesson is that the loop-current phase is promoted by the same interaction as the nematic phase, $V_{pp}$. At the same time, the loop-current phase does not depend on a high density of states at the Fermi level. Hence a  Fermi surface deformation due to nematic instability has no direct influence on loop-current instability, and the two can therefore coexist. The possible coexistence of nematic and loop-current phases we find in
this work is interesting in light of experimental observations of
both IUC nematic order\cite{hinkov:2008, daou:2010} and  IUC staggered
magnetism in underdoped YBCO\cite{fauque:2006}.

\section{Closing Remarks}
\label{sec:closing}
In this article, we reviewed recent progress from our``middle-up/down'' approach to ELC phenomena in underdoped cuprates. In section~\ref{sec:nematic} we discussed the introduction and detection of the IUC nematic order parameter for the pseudogap states of cuprate superconductors~\cite{Lawler:2010fk}, which motivated much of the rest of the article. 
In section~\ref{sec:coupling} we discussed a 
phenomenological theory of the coupling between nematic and smectic order parameters and its application to stripe dislocations\cite{Mesaros22072011}, building on the 
 analysis scheme developed for Ref.~\cite{Lawler:2010fk}. 
In section~\ref{sec:emery} we discussed microscopic mechanisms for various IUC symmetry breaking\cite{PhysRevB.84.144502}, building on 
the importance of oxygen sites in IUC nematic observation in Ref.~\cite{Lawler:2010fk}. These discussions amount to a strong experimental and theoretical support for (local) ELC formation in the underdoped cuprates in so-called ``pseudogap regime'' (see Fig~\ref{fig:pd}). In particular, by providing direct evidence of ELC formation in \BSCCO, 
 they establish ELC formation as a universal aspect of underdoped cuprates\footnote{In La-based systems, stripe formation is well established through neutron scattering~\cite{Tranquada:2004fk}. In YBCO,  transport anisotropy~\cite{Ando:2002prl}and neutron scattering~\cite{hinkov:2008} showed evidence of nematic. More recently, in-field NMR show strong evidence of a field stablized charged stripe~\cite{Wu:2011uq}}.

Given evidence of broken symmetries associated with ELC,  it is natural to shift focus to the nature of the symmetry-broken phases and their relation to superconductivity. Two key questions are (i) how do ELC phases form and (ii) what are their relations to spectral gap and superconductivity.

First regarding how ELC phases form, theoretical models for smectic(stripe) formation and those for nematic formation have been discussed separately. Existing models for stripe formation include: Hartree-Fock approximations that find insulating stripes driven by a reduction in the kinetic energy of the holes that move transverse to the stripe\cite{PhysRevB.39.9749, PhysRevB.40.7391,Machida1989192,PhysRevLett.64.1445}, Coulomb frustrated phase separation approach building on the role of long-range Coulomb interactions\cite{Emery1993597}, and DMRG calculations on t-J models that find d-wave pairing correlations in charge stripes (see the review article by Scalapino and White in this issue for more details.). 
As for the nematic formation though it has been conjectured that a stripe phase would {\em melt} into a nematic phase through quantum or thermal fluctuations(see Fig.~\ref{fig:strong-weak}), there are only phenomenological models for such transitions\cite{sun-2008,PhysRevLett.104.106405}. Instead, alternative approaches ignore possible stripe formation and instead begin from a Pomeranchuk instability~\cite{pomeranchuk-1958} and assume an attractive forward-scattering interaction in the $d-$wave channel that causes a transition into the nematic phase from an isotropic fermi-liquid phase~\cite{oganesyan:2001,kee:2003}. To place these theories in a more concrete context, the mean-field theory of Emery model reviewed in section~\ref{sec:emery} showed an on-site $U_d$ repulsion and an inter-oxygen $V_{pp}$ repulsion can generate such an effective attractive interaction.

However, observations of the coexistence of nematic order and fluctuating stripes\cite{Lawler:2010fk, PhysRevLett.104.106405, hinkov:2008, Daou:2010} are calling for microscopically motivated models that can treat both nematic and smectic formation. Recent phenomenological models offer new insight into what elements a successful such microscopic model should capture. Ref.~\cite{PhysRevB.84.012507} demonstrated
coexisting charge stripe and nematic order could explain both NMR\cite{Wu:2011uq} and quantum oscillations 
on YBCO under a high magnetic field that suppresses superconductivity first discovered in Ref.~\cite{quantum-oscillation}.  
Ref.~\cite{2012arXiv1201.3358M} showed that a tractable theory of quantum melting of stripes is possible when such melting is through proliferation of double dislocations, which could be preferred over single dislocations in the presence of an antiferromagnetic background. Given the lessons on nematic formation from Ref.~\cite{PhysRevB.84.144502}, progress in capturing antiferromagnetism, stripe and nematic might be possible starting from the Emery model. 

Now we turn to the second key open question: relations between ELC and spectral gap or superconductivity. While insulating fully filled stripe solution from Hartree-Fock studies~\cite{PhysRevB.39.9749, PhysRevB.40.7391,Machida1989192,PhysRevLett.64.1445} only show how stripe formation would inhibit superconductivity, DMRG calculations on the t-J model show that $d-$wave pairing correlations make the stripes half-filled and more consistent with experiments (see the article by Scalapino and White in this issue). These DMRG observations make a concrete case that superconductivity and stripe formation might not be simply mutually exclusive. When it comes to the relationship between nematic order and superconductivity, much less is known. Ref.~\cite{PhysRevB.77.184514} showed that nematic quantum phase transition can be a continuous transition inside the superconducting dome, through a shift of nodal positions. This is only a proof that nematic and superconductivity can coexist, but not a proof that nematic order would help or hurt superconductivity. Though many experiments show nematic phenomena in the psuedogap phase, as of now there is no theoretical evidence of a nematic phase causing a spectral gap. However, this could have been due to shortage of studies of strongly correlated microscopic models with nematic ordered ground states. It is quite possible the underlying antiferromagnetic fluctuations, absent in many approaches, are needed to pin down why nematic phenomena is seen in the pseudogap phase.

\appendix
\setcounter{figure}{0}
 \section{Correcting Slow Drift in Topography and Electronic Structure}
\begin{figure}[tb]
\centering
    \subfigure[]{\includegraphics[width=.47\textwidth]{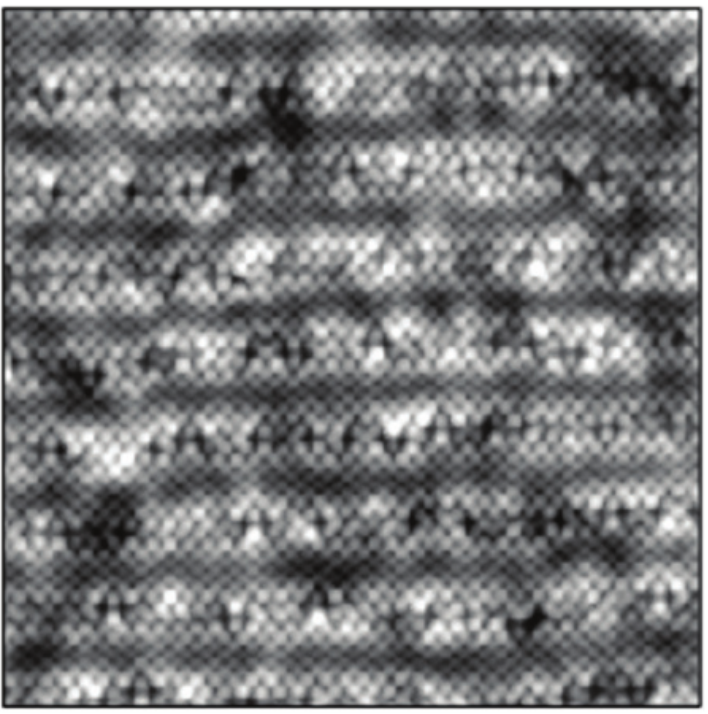}}
   \subfigure[]{ \includegraphics[width=.47\textwidth]{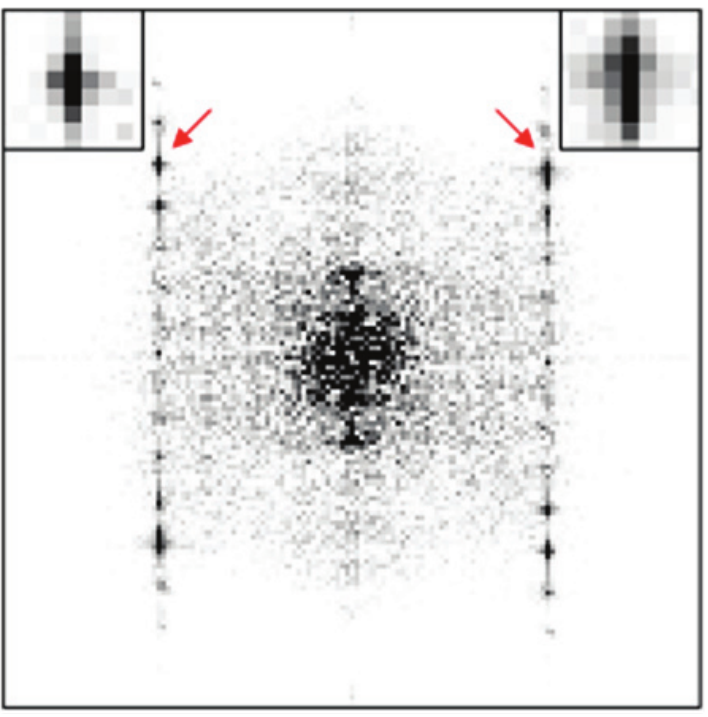}}
  \caption{An example Topograph taken from the supplementary materials of Ref.~\cite{Lawler:2010fk}. (a) the real space image showing both lattice modulations and longer wavelength horizontal superlattice modulations. (b) Fourier transform of (a) showing lattice Bragg peaks (highlighted with arrows), superlattice generated harmonics and a broad longer wave length structure.}
  \label{fig:Topo}
\end{figure}

Here we describe how to use the topograph to correct for the picometer scale drift of the tip location (typically due to both piezoelectric mechanical creep and small temperature variations during data acquisition). 
A typical topograph, as shown in Fig. \ref{fig:Topo}., shows variations at differing length scales associated with different physics: modulations with wave vectors  $Q_x$ and $Q_y$  for the Bi (and Cu) lattice sites, a super-lattice modulation with wave vector $Q_{sup}$, and the mentioned slowly varying apparent “displacement” due to long-term and picometer scale piezoelectric drift. The latter is an artifact of the experimental process and not part of the physical measurement and is best removed when possible. We define the slowly varying “displacement” field $\vec u(\vec r)$ such that un-displaced positions  $\vec r - \vec u(\vec r)$ (which are the Lagrangian coordinates of elasticity theory) will form a perfect square lattice with Cu lattice locations $\vec d_{Cu}=0$. Now the topograph is expected to take the form
\begin{equation}
  T(\vec r) = T_0\left(\!\cos\vec Q_x\!\cdot\!\big(\vec r\!-\!\vec u(\vec r)\big)\! +\! \cos\vec Q_y\!\cdot\!\big(\vec r\!-\!\vec u(\vec r)\big)\!\right) + T_{sup}\cos\vec Q_{sup}\!\cdot\!\big(\vec r\!-\!\vec u(\vec r)\big) + \ldots
\end{equation}
where $\ldots$ represents other contributions such as impurities etc. That $\vec u(\vec r)$ is slowly varying compared to the scale of the super-lattice modulation and the lattice modulations is evident from the Fourier transform of the topograph. In order to extract the slow varying $\vec u(\vec r)$, it is useful to introduce a coarsening length scale $1/\Lambda_u$ over which $\vec u(\vec r)$ is roughly constant such that $\Lambda_u \ll \rm{Min}(|\vec Q_{sup}|,|\vec Q_{x,y}|)$. The Fourier transform of the topograph shows that we can quite safely choose a small $\Lambda_u$ since the lattice peak is sharp.  Now consider
\begin{equation}
  T_x(\vec r) = \sum_{\vec r'} T(\vec r')e^{-i\vec Q_x\cdot\vec r'}\bigg(\frac{\Lambda_u^2}{2\pi}e^{-\Lambda_u^2|\vec r-\vec r'|^2/2}\bigg)
\end{equation}
the weighted average of $T(\vec r)e^{-i\vec Q_x\cdot\vec r}$ over the length scale $1/\Lambda_u$. Since $\Lambda_u\ll|Q_{sup}|,|Q_y|$, their contributions average out, leaving
\begin{equation}
  T_x(\vec r) \approx (T_0/2)e^{-iQ_x\cdot\vec u(\vec r)}
\end{equation}
Here, we made use of the fact that $\vec u(\vec r')\approx \vec u(\vec r)$   for small $|\vec r-\vec r'|\ll 1/\Lambda_u$. We can define the y-component $Q_y\cdot\vec u(\vec r)$ in a similar fashion. Hence we can extract the full displacement field $\vec u(\vec r)$ (much as in the spirit of elasticity theory) and thus undo all effects of both piezoelectric drift and also set the origin of the coordinate system such that $\vec d_{Cu}=0$.

It turns out that the drift over the extent of a typical image is approximately one or two lattice-spacings out of $>100$ or less than 10 picometer per unit cell. This manifests itself through the phases $\vec Q_x\cdot\vec u(\vec r)$  and $\vec Q_y\cdot\vec u(\vec r)$ which jump by 2π in some regions. These phase jumps need to be removed to make $u(\vec r)$ a single valued quantity. By taking a derivative of the image to locate the jumps and adding $2\pi$ to $\vec u(\vec r)$ where appropriate can perform this function. Finally, and importantly, the same geometrical transformations that define $T_x$ and $T_y$ are carried out on each $Z(\vec{r},e)$ acquired simultaneously with the $T(\vec r)$ so that both are registered to each other while being rendered periodic with the lattice; without these procedures, the $O_N(e)$ function of the main text cannot be evaluated correctly.

\section*{Acknowledgements}
E-AK acknowledges the support from the NSF
Grant DMR-0520404
 to the Cornell Center for Materials Research,  
and from the NSF CAREER grant DMR-0955822. 









\end{document}